\documentclass[aps,prd,twocolumn,showpacs,superscriptaddress,dblfloatfix,preprintnumbers]{revtex4-1}
\usepackage{graphicx}   
\usepackage{dcolumn}   
\usepackage{bm}           
\usepackage{amssymb}  
\usepackage{amsmath}
\usepackage{dsfont}
\usepackage{setspace}
\usepackage{amsmath, amssymb, setspace}
\usepackage{array}
\usepackage{booktabs}
\usepackage{mathrsfs}
\usepackage{indentfirst}
\usepackage{slashed}
\usepackage{float}
\usepackage{lmodern}
\usepackage{hyperref}
\usepackage{xcolor}
\usepackage{multirow}
\usepackage{epstopdf}
\usepackage{ulem}
\usepackage{tabularx}
\usepackage[flushleft]{threeparttable}
\usepackage{hyperref}
\usepackage{soul}
\usepackage{braket}
\usepackage[T1]{fontenc} 
\usepackage[utf8]{inputenc} 

\DeclareUnicodeCharacter{246}{\"o}	
\DeclareUnicodeCharacter{350}{\c{S}}

\begin{document}

\newcommand{\eV}{\text{eV}}
\newcommand{\keV}{\text{keV}}
\newcommand{\MeV}{\text{MeV}}
\newcommand{\GeV}{\text{GeV}}
\newcommand{\arcmin}{\text{arcmin}}
\newcommand{\Mpc}{\text{Mpc}}
\newcommand{\Hunit}{$\text{km\,s}^{-1}\,\text{Mpc}^{-1}$}
\newcommand{\muK}{$\mu\text{K}$}
\newcommand{\beq}{\begin{equation}}
\newcommand{\eeq}{\end{equation}}
\newcommand{\bea}{\begin{eqnarray}}
\newcommand{\eea}{\end{eqnarray}}
\newcommand{\dd}{\partial}
\newcommand{\greeksym}[1]{{\usefont{U}{psy}{m}{n}#1}}
\newcommand{\umu}{\mbox{\greeksym{m}}}
\newcommand{\udelta}{\mbox{\greeksym{d}}}
\newcommand{\uDelta}{\mbox{\greeksym{D}}}
\newcommand{\uPi}{\mbox{\greeksym{P}}}
\newcommand{\F}{\Phi}
\newcommand{\f}{\phi}
\newcommand{\vf}{\varphi}
\newcommand{\Q}{\tilde{Q}_{_L}}
\newcommand{\q}{\tilde{q}_{_R}}
\newcommand{\Lp}{\tilde{L}_{_L}}
\newcommand{\lp}{\tilde{l}_{_R}}
\newcommand{\nL}{$\, \backslash \! \! \! \! L \ $}
\newcommand{\mh}{m_{_{\rm H}}}
\newcommand{\ms}{m_{_{\rm S}}}
\newcommand{\lh}{\lambda_{_{\rm H}}}
\newcommand{\ls}{\lambda_{_{\rm S}}}
\newcommand{\lm}{\lambda_{_{\rm HS}}}
%

\newcommand{\D}{\text{D}}
\newcommand{\ud}{\text{d}}
\newcommand{\curl}{\,\text{curl}\,}


\newcommand{\Omtot}{\Omega_{\mathrm{tot}}}
\newcommand{\Omb}{\Omega_{\mathrm{b}}}
\newcommand{\Omc}{\Omega_{\mathrm{c}}}
\newcommand{\Omm}{\Omega_{\mathrm{m}}}
\newcommand{\omb}{\omega_{\mathrm{b}}}
\newcommand{\omc}{\omega_{\mathrm{c}}}
\newcommand{\omm}{\omega_{\mathrm{m}}}
\newcommand{\omnu}{\omega_{\nu}}
\newcommand{\Omnu}{\Omega_{\nu}}
\newcommand{\Oml}{\Omega_\Lambda}
\newcommand{\OmK}{\Omega_K}




\newcommand{\sigt}{\sigma_{\mbox{\scriptsize T}}}


\preprint{UCI-TR-2019-20, arXiv:1907.11696}

\title{\boldmath Hidden Treasures: sterile neutrinos as dark matter with miraculous abundance, structure formation for different production mechanisms, \\ and a solution to the \texorpdfstring{$\sigma_8$}{sigma-8} problem }

\author{Kevork N.\ Abazajian}
\affiliation{Department of Physics and Astronomy, University of California, 
Irvine, CA 92697, USA}

\author{Alexander Kusenko}
\affiliation{Department of Physics and Astronomy, University of California, Los Angeles,
 CA 90095-1547, USA}
\affiliation{Kavli Institute for the Physics and Mathematics of the Universe (WPI), UTIAS, 
The University of Tokyo, Kashiwa, Chiba 277-8583, Japan}

\date{\today}

\begin{abstract}
We discuss numerous mechanisms for production of  sterile neutrinos, which can account for all or a fraction of dark matter, and which can range from warm to effectively cold dark matter, depending on the cosmological scenario. We investigate production by Higgs boson decay, $(B-L)$ gauge boson production at high temperature, as well as  production via resonant and nonresonant neutrino oscillations.  We calculate the effects on structure formation in  these models, some for the first time.  If two populations of sterile neutrinos, one warm and one cold, were produced by different mechanisms, or if sterile neutrinos account for only a fraction of dark matter, while the remainder is some other cold dark matter particle, the resulting multi-component dark matter  may alleviate some problems in galaxy formation.  We examine the X-ray constraints and the candidate signal at 3.5 keV.  Finally, we also show that the $\sigma_8$ problem can be a signature of fractional dark matter in the form of sterile neutrinos in several mechanisms.

\end{abstract}


\maketitle

\section{Introduction}
  
Sterile or right-handed neutrinos are introduced for the purpose of explaining the observed masses of active neutrinos.  Since the observed neutrino masses depend only on the ratio of the unknown Yukawa coupling to the mass of the right-handed neutrino, that right-handed neutrino's Majorana mass has an enormous range of allowed values, from the eV to the Plank scale.  Naturalness arguments can be made in favor of both large and small Majorana neutrino masses~\cite{Kusenko:2009up}.  In the large mass limit, the right-handed neutrinos have no effect on the low-energy effective theory (although they could play an important role in cosmology by generating the  matter-antimatter asymmetry of the universe). However, if one of the Majorana masses is of the order of 1-10~keV, the corresponding particle can be dark matter~\cite{Dodelson:1993je,Abazajian:2001nj} and can affect  supernova explosions in ways consistent with observations~\cite{Kusenko:1997sp,Fuller:2003gy}. This dark matter candidate arises from a very minimal extension of the Standard Model by one light sterile neutrino. A model with three sterile neutrinos below the electroweak scale dubbed $\nu$MSM~\cite{Asaka:2005an,Asaka:2005pn} has been widely discussed in connection with dark matter and leptogenesis. 

The dark matter population of sterile neutrinos could be produced by the oscillations of active neutrinos into sterile or by some other mechanism. If the neutrino oscillations are responsible for the entire population of relic sterile neutrinos, the dark matter particles are produced at temperatures below a GeV, and they constitute warm dark matter (WDM).  Furthermore, since the same mixing parameter controls the production and the decay of sterile neutrinos in the case of oscillation production, the X-ray signatures expected from dark matter are uniquely determined by the particle mass and the mixing that produce the requisite abundance of sterile neutrino dark matter~\cite{Abazajian:2001nj,Abazajian:2001vt}. 

Alternatively, a population of dark mater in the form of sterile neutrinos can be produced by another mechanism.  If the mixing parameters are small and neutrino oscillations are not efficient enough to generate the full dark matter abundance, some or most of dark matter can be made up of the sterile neutrinos with some very different free-streaming properties. The change in the number of degrees of freedom due to the QCD transition results in dilution and redshifting of any out-of-equilibrium  population produced at temperatures higher than a GeV. So, if sterile neutrinos are produced at a higher temperature, they constitute a much colder form of dark matter.  Furthermore, the expected X-ray signatures can be suppressed by the small mixing angle, while the abundance of sterile neutrinos can still be large enough to account for all dark matter.  We will consider several such high-scale scenarios and identify their predictions for the dark matter properties. We also calculate, for the first time, the linear transfer functions for several production mechanisms---Higgs decay and two types of GUT-scale production---to assess their effects on cosmological structure formation, as they cross the regime from cold to warm dark matter. 

\section{The keV miracle model: Higgs decay}
The natural abundance of sterile neutrinos produced in singlet Higgs boson decays is an appealing feature of a freeze-in production scenario at the electroweak scale~\cite{Kusenko:2006rh,Petraki:2007gq,Kusenko:2009up,Kang:2014cia}.  If the Majorana mass arises from the Higgs mechanism, the corresponding Higgs boson must be a singlet with respect to the standard model gauge group.  Assuming that the singlet Higgs $S$ has mass and VEV of the order of the electroweak scale, $\langle S \rangle \sim m_S\sim 10^2{\rm\, GeV}$, and as long as the dark-matter sterile neutrino mass is in the 1-10 keV range (which is necessary for dark matter) the dark matter abundance comes out to be correct. 

Let us recap the essential elements of this model~\cite{Kusenko:2006rh,Petraki:2007gq}. We consider the following Lagrangian: %
\begin{align} 
{\mathcal L} 
   = &  {\mathcal L}_{0}+\bar N_{a} \left(i \gamma^\mu \partial_\mu 
\right ) N_{a}  \nonumber \\
 - & y_{\alpha a} H \,  \bar L_\alpha N_{a}  - \frac{f_a}{2} \, S \,
\bar {N}_{a}^c N_{a} 
 +V(H,S) + h.c. \,, 
\label{lagrangianS}
\end{align}
where $ {\mathcal L}_{0}$ includes the gauge and kinetic terms of the Standard
Model, $H$ is the Higgs doublet, $S$ is the real boson which is SU(2)-singlet,
$L_\alpha$ ($\alpha=e,\mu,\tau$) are the lepton doublets, and $ N_{a}$
($a=1,...,n$) are the additional singlet neutrinos.  
 
The most general renormalizable scalar potential consistent with the symmetries has the form: 
\begin{align}
V(H,S) &=   \mu_{H}^2 |H|^2 + m_{2}^2 S^2  + \lambda_3 S^3 \nonumber \\  & +  
\lambda_{_{HS}} |H|^2 S^2+ \lambda_{_S}  S^4   + \lambda_{_H}
|H|^4 ,
\label{potential}
\end{align}
\beq
\langle H\rangle= v_0=247 \ {\rm GeV}, \ \langle S\rangle= v_1\sim v_0,
\eeq
and the singlet fermions acquire the Majorana masses  $m_a = f_a v_1$. In this model, the only source of the Majorana masses is the Higgs mechanism (via a gauge singlet Higgs boson), while the tree level Majorana mass can be forbidden by a discrete symmetry. 

The dark matter particle is $N_1$, and the Yukawa coupling is chosen so that 
\beq
m_1=f_1 \langle S\rangle \sim {\rm keV} \ \Rightarrow \ f_1\sim 10^{-8}.
\eeq
One can easily check that, for the Yukawa coupling as small as $f_1\sim 10^{-8}$, the $N_1$ particles do not come to equilibrium at any temperature. 
The mixing terms in the scalar potential can guarantee that the $S$ particles are in thermal equilibrium at temperatures above $m_S$. 

The presence of the $SNN$ term has an important consequence:  in addition to generating the Majorana masses, this term opens a new production channel for $N_1$ particles via decays $S\rightarrow NN $, while the $S$ particles in equilibrium.   At later times the sterile neutrinos remain out of equilibrium, while their density and their momenta get red-shifted by the expansion of the universe making dark matter ``colder.''  One can estimate the number density $n_s$ of  sterile neutrinos by multiplying the $S$ number density (which is $\sim T^3$ for $T>m_S$)  by the $S \rightarrow NN$ decay rate,  $\Gamma_S = ({f^2}/{16\pi})m_S$ and the time available for decay, $\tau \sim M_0/T^2,$ at the lowest temperature when the $S$ particles are in equilibrium, $T\sim m_S$.  The result is 
\beq
 \left( \left. \frac{n_s}{T^3}  \right) \right |_{T\sim m_S}\sim \Gamma 
\left.  \frac{ M_0}{T^2}\right |_{T\sim m_S} \sim \frac{f^2}{16\pi}
\frac{M_0}{m_S},
\label{approx_n_s}
\eeq
where $M_0 = ({45 M_{PL}^2}/{4 \pi^3 g_*})^{1/2}  \sim 10^{18}\, \GeV$
is the reduced Planck Mass.  This approximate result is in good agreement with a more detailed calculation~\cite{Shaposhnikov:2006xi,Petraki:2007gq}. The mass density is obtained by multiplying $n_s$ by the dark matter particle mass, $f_1 \langle S \rangle$: 
\beq
 \left( \left. \frac{\rho_s}{T^3}  \right) \right |_{T\sim m_S}  \sim \frac{f^3}{16\pi}
\frac{M_0 \langle S \rangle}{m_S}.
\label{approx_rho_s}
\eeq

Once produced, the dark-matter particles remain out of equilibrium.  The entropy production at the QCD transition temperature dilutes the 
density by some factor $\xi$. Assuming only the degrees of freedom in the  
Lagrangian of Eq.~(\ref{lagrangianS}), that is, the Standard Model with the addition of $N$ and $S$ fields, one
obtains $\xi =g_*(T=100\, {\rm GeV})/g_*(T=0.1\, {\rm MeV})\approx 33.$
Therefore, the density of dark matter is given by 
\beq
 \left( \left. \frac{\rho_s}{T^3}  \right) \right |_{T< {\rm MeV}}  
 \sim \frac{1}{\xi} \frac{f^3}{16\pi }
\frac{M_0 \langle S \rangle}{m_S}
=\frac{m_1^3 M_0}{16\pi \xi m_S \langle S \rangle^2}
\sim {\rm eV} ,
\label{approx_rho_DM}
\eeq
that is, exactly the observed present value of $\rho_{\rm DM}/T_\gamma^3$, which corresponds to $\Omega_{\rm DM}=0.2$.  This coincidence of scales to produce the proper dark matter density is unique among the models for sterile neutrino dark matter production, and it can be compared with the ``WIMP miracle'' of electroweak-scale dark matter.  For this reason we dub this model the {\it keV Miracle Model}. 

\section{Production at the Grand Unified Theory scale}

The Split Seesaw model~\cite{Kusenko:2010ik} produces two large and one small Majorana masses due to a natural separation of scales.  The large Majorana masses allow for thermal leptogenesis, while the small, keV mass produces a dark matter candidate.  The model can be embedded into an SO(10) Grand Unified Theory, or some other theory containing a gauge U(1)$_{\rm B-L}$ symmetry.  Sterile neutrinos couple to the U(1)$_{\rm B-L}$ boson, which opens two scenarios for dark matter production in this model.  

\subsection{GUT Scenario 1}
 If the reheat temperature is high enough to restore the U(1)$_{\rm B-L}$ symmetry, sterile neutrinos reach thermal equilibrium through interactions with U(1)$_{\rm B-L}$ bosons.  As the temperature of the universe decreases, the gauge U(1)$_{\rm B-L}$ symmetry must be broken.  The corresponding phase transition can be of the first order, leading to a significant entropy production.  In the broken phase, the $(B-L)$ gauge boson is massive, and sterile neutrino is out of equilibrium.  Of course, if the density of sterile neutrinos remained equal to their thermal density in the symmetric vacuum, their abundance would be higher than needed for dark matter.  However, the entropy released in the phase transition can dilute this density by factor $\xi\sim O(10^2)$ to the value consistent with the observations. There is a broad range of parameters for which this can be realized~\cite{Kusenko:2010ik}. At the same time, the momenta of the dark matter particles are red-shifted by the factor $\xi^{1/3}$, similar to the keV Miracle Model described above.

\subsection{GUT Scenario 2}

An alternative production scenario assumes that the reheating temperature, $T_R$ is below the U(1)$_{\rm B-L}$ symmetry breaking temperature.  In this case, the density of the sterile neutrinos never reaches the thermal density.  To obtain the correct dark matter abundance, the reheating temperature must have a specific value, which turns out to be a (reasonable) $T_R\sim 5 \times 10^{13} \, {\rm GeV}$~\cite{Kusenko:2010ik}.  In this case, the distribution of sterile neutrinos at the time of production is closer to the thermal distribution with the temperature $T_R$ as compared to the Scenario 1 described above.  The low-energy transfer function is affected by the entropy production just as in the keV Miracle Model, and the resulting average momentum is close to that of the Miracle Model.  

If there are particle thresholds between the Weak scale and the GUT scale, the resulting change in the number of degrees of freedom can lead to entropy production and further cooling of dark matter in Scenarios 1 and 2.

\section{Nonresonant and Resonant Oscillation Production}

The first mechanism proposed for production of sterile neutrino dark matter was that of a standard baryon number $B$ and lepton number $L$ symmetric, $B-L=0$, thermal history where production proceeds through neutrino oscillations. The effects of the relaxation of suppression of active-sterile mixing due to the lowering of the neutrinos' thermal potential allows for production through the Dodelson-Widrow mechanism, where collisions produce sterile neutrinos from intervening active-sterile oscillations \cite{Dodelson:1993je}. The production is predominantly at temperatures of $T\sim 133{\rm\, MeV}(m_s/\rm keV)^{1/3}$, and the proper cosmological dark matter density is achieved by matching the mixing angle, described by the mass-mixing production relation 
\begin{equation}
    m_s = 3.4{\rm\, keV}\left(\frac{\sin^2 2\theta}{10^{-8}}\right)^{-0.615}\left(\frac{\Omega_s}{0.26}\right)^{1/2}\, , \label{DWProduction}
\end{equation}
for a standard quark-hadron cross-over transition, and production of a fraction of critical density of $\Omega_s$ \cite{Abazajian:2005gj}. Note, of course, that $\Omega_s$ can be less than $\Omega_\mathrm{DM}$, allowing for a fraction of dark matter to be sterile neutrinos. 

The average momentum to temperature of this model is typically $\langle p/T \rangle\approx 2.8 $. This model has been ruled out as being responsible for all of the cosmological dark matter through a combination of Local Group galaxy counts and X-ray flux limits. The latter constraints come from the fact that the sterile neutrino would radiatively decay and can be obtained once the mass and mixing are fixed to produce the observed total dark matter density \cite{Horiuchi:2013noa}. We discuss this mechanism as a potential source of a partial fraction of the full dark matter density, with the rest of the dark matter due to another sterile neutrino production mechanism, of the same sterile neutrino or another sterile state, or a less related dark matter particle. The fraction of dark matter as sterile neutrinos allows for a few interesting phenomena: first, all other production models will have a minimal level from oscillation production given by Eq.~\eqref{DWProduction}; second, a fraction of $\sim$15\% of the dark matter to be produced by the Dodelson-Widrow mechanism can be responsible for the 3.55 keV X-ray line detected in several observations \cite{Bulbul:2014sua,Boyarsky:2014jta}, and with the remainder of the dark matter being cold, this model can escape structure formation bounds \cite{Diamanti:2017xfo,Abazajian:2017tcc}; and third, a fraction of sterile neutrinos as dark matter in the Dodelson-Widrow mechanism can be responsible for alleviating the $\sigma_8$ problem, as discussed in Section \ref{FractionSection}.

Work by Shi \& Fuller \cite{Shi:1998km} pointed out that a nonzero lepton number universe ($B-L\neq 0$) creates a matter potential that can produce a Mikheev-Smirnov-Wolfenstein (MSW) resonance, enhance production for smaller mixing angles, and provide a cooler average $\langle p/T\rangle$ than nonresonant production. The range of production of the Shi-Fuller mechanism has been calculated in greater detail by Venumadhav et al. \cite{Venumadhav:2015pla}. It has been shown that the Shi-Fuller mechanism can be responsible for the 3.55 keV line, and potentially alleviate issues with structure in the Local Group of galaxies \cite{Abazajian:2014gza}. However, there remains tension between Local Group satellite counts and X-ray limits for this model \cite{Cherry:2017dwu}. We will discuss fractional production of this model in \S\ref{FractionSection}.

\begin{table}[t!]
\centering
\begin{tabular}{l|c|c}
model &	$\langle p/T\rangle$	& References \\
\hline
Dodelson-Widrow	& 2.83 & \cite{Abazajian:2005gj} \\
Shi-Fuller	& 1.3 to 2.6 & \cite{Venumadhav:2015pla}\\
keV Miracle Model  	& 0.76 & \cite{Petraki:2007gq}\\
GUT scale scenario 1 &	0.2 & \cite{Kusenko:2010ik}\\
GUT scale scenario 2 &	0.7 &  \cite{Kusenko:2010ik}
\end{tabular}
\caption{\label{table:pt} We tabulate a summary of the models discussed in the text with their respective average momenta per temperature for the distributions arising out of their production, with relevant references. }
\end{table}

\section{Cosmological Structure Formation}
\subsection{The keV Miracle Model: Higgs Decay}
\begin{figure}
    \centering
    \includegraphics[width=3.4in]{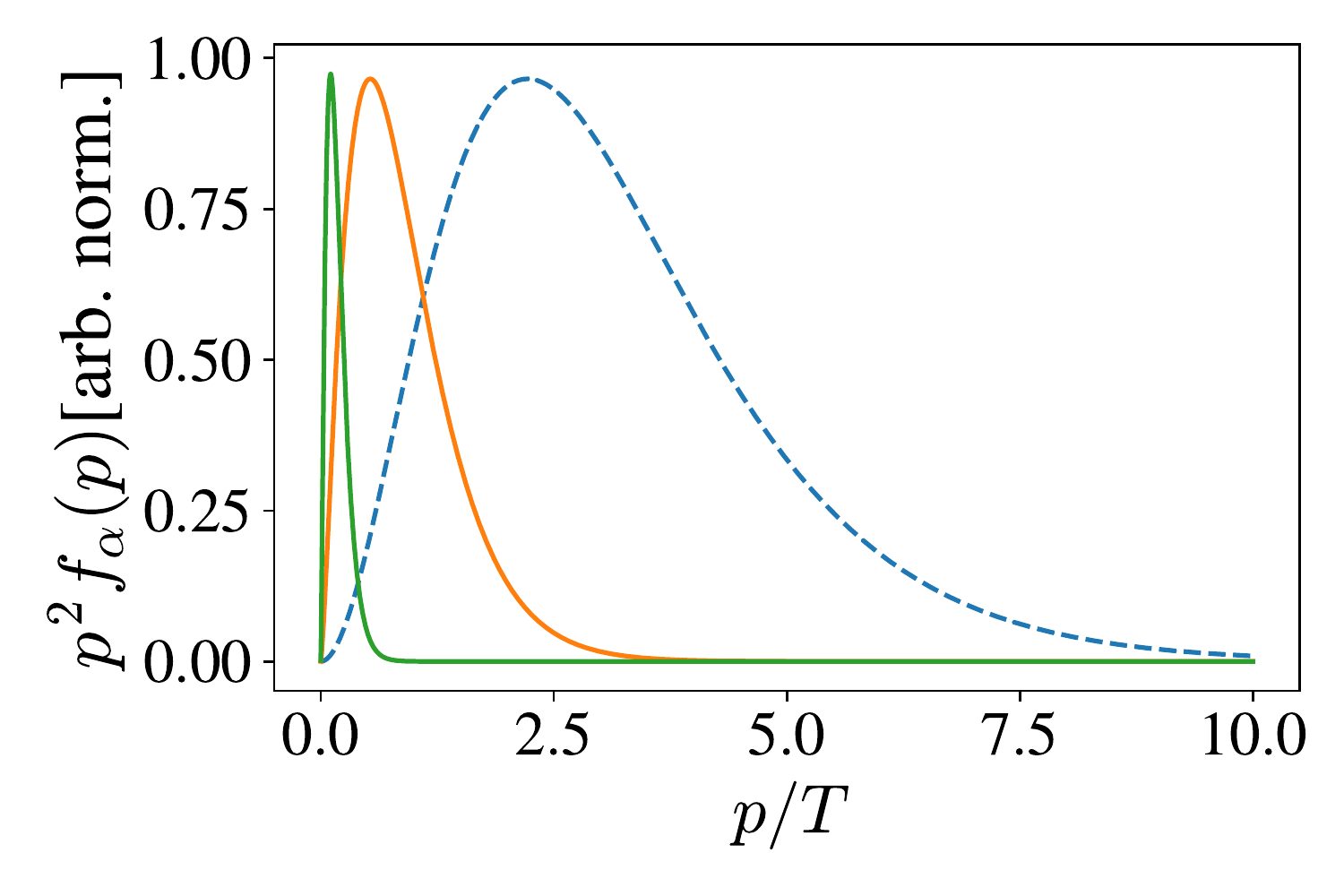}
    \caption{The phase-space distributions of the keV Miracle Model (Higgs decay, orange) and $B-L$ high-temperature boson decay model (green) relative to thermal (dashed blue).}
    \label{fig:dist}
\end{figure}

In this model, the Higgs singlet $S$ are in thermal equilibrium, and provide an energy distribution to the sterile neutrinos $N_1$ that is non-thermal, and cooled due to the disappearance of degrees of freedom between production and the onset of structure formation. In this case, the momentum-energy distribution, $f$, of the sterile neutrinos is \cite{Petraki:2007gq}
\beq
f(x) \propto x^2\int_1^\infty\frac{(z-1)^{3/2}}{e^{xz}-1}\, dz\, ,
\eeq
where $x\equiv p/T$, and the normalization is given by the cosmological dark matter density. To first order, the effects on structure formation can be ascertained by the average momentum relative to the temperature of the plasma. For this case, the distribution goes from that immediately after production of $\langle p\rangle/T|_{100\,\mathrm{GeV}} \approx 2.45$ to a cooler one after disappearance of degrees of freedom in the plasma including the Standard Model particles and those in the model, so that $\langle p\rangle/T|_{\ll 1\,\mathrm{MeV}} \approx 0.76$ (while a thermal distribution has $\langle p\rangle/T \approx 3.15$. The distribution function is plotted in Fig.~\ref{fig:dist}.

\begin{figure}
    \centering
    \includegraphics[width=3.4in]{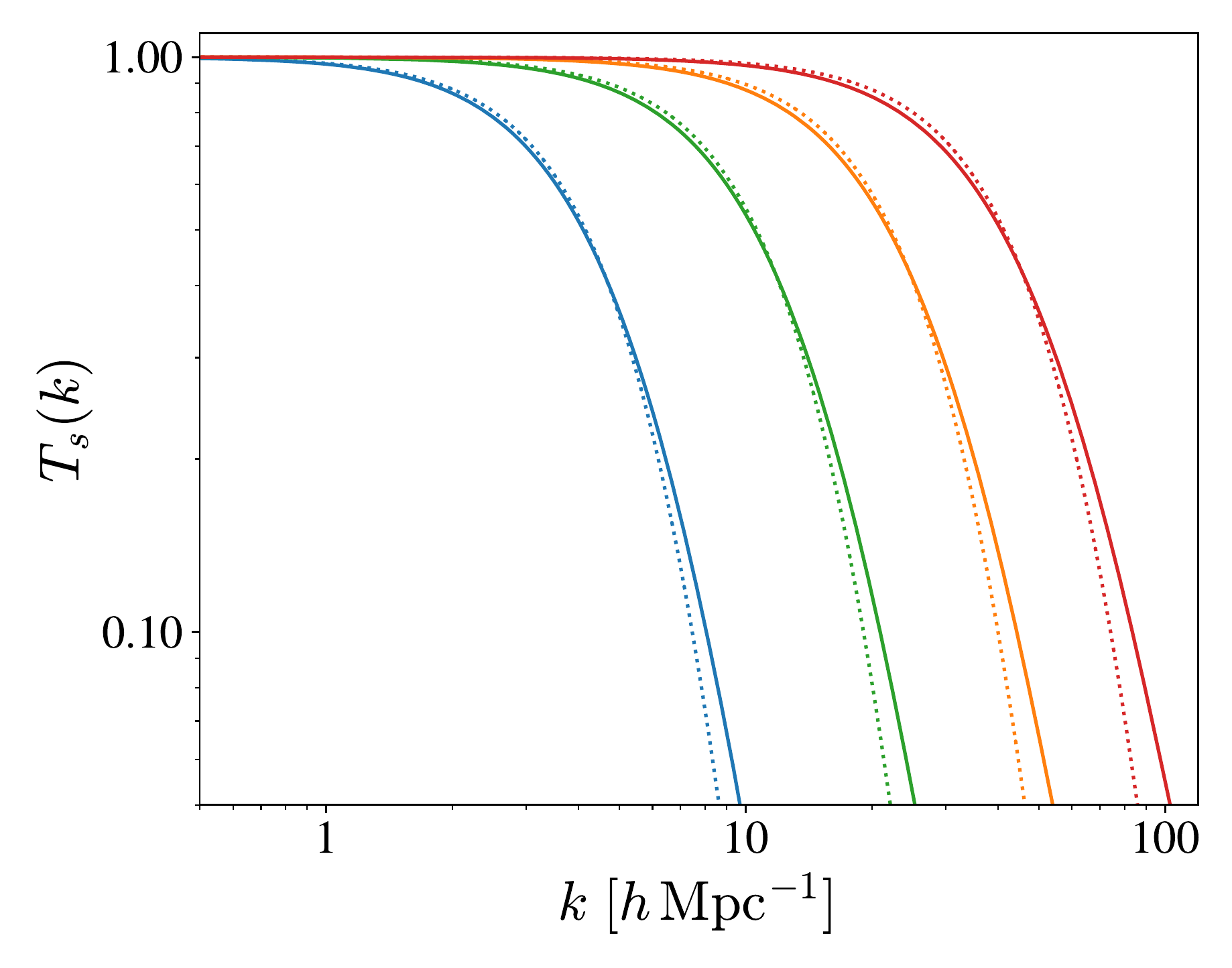}
    \caption{The transfer functions of sterile neutrino dark matter in the case of the sterile neutrino keV Miracle Model (Higgs decay). The transfer functions $T_s(k)$ are shown for increasing cut-off wave number $k$ with increasing mass $m_s = 1, 3, 7,\text{and}\ 14\,\mathrm{keV}$. The equivalent thermal WDM particle mass equivalent transfer functions are shown in dashed lines, and correspond to $m_\mathrm{WDM} =  0.45, 1.0, 2.0, 3.3 \rm\, keV $, respectively.}
    \label{fig:PK}
\end{figure}

In order to quantify more precisely the effects of this model on structure formation, we modify the cosmological Boltzmann code CAMB \cite{Lewis:1999bs} to include a modified sterile neutrino energy distribution function \cite{Abazajian:2005gj}. For sterile neutrino masses of $m_s = 1, 3, 7,\text{and}\ 14\,\mathrm{keV}$ we include the production momentum distributions in the full Boltzmann transport as calculated by CAMB. The linear clustering of the matter power spectrum relative to pure cold dark matter is parametrized by the sterile neutrino transfer function
\begin{equation}
    T_s(k) \equiv \sqrt{\frac{P_\mathrm{sterile}(k)}{P_\mathrm{CDM}(k)}}\, ,
\end{equation}
where $P_\mathrm{sterile}(k)$ and $P_\mathrm{CDM}(k)$ are the linear matter power spectra for the pure sterile neutrino dark matter model and CDM, respectively. The transfer functions are shown in Fig.~\ref{fig:PK}. Due to the ``cool'' nature of the miracle distributions, the cutoff scales for structure growth in this model are smaller than one obtains in oscillation based production in standard cosmologies (the Dodelson-Widrow model \cite{Dodelson:1993je}), with transfer functions that are well approximated by transfer functions of thermal WDM with particle masses of 0.45, 1.0, 2.0 and 3.3 keV for the $m_s = 1, 3, 7,\text{and}\ 14\,\mathrm{keV}$ cases, respectively. We calculate the thermal WDM particle mass equivalent fit and relation for thermal WDM transfer functions as Eq.~(A8) in Ref.~\cite{Bode:2000gq}. One could also fit to the more generalized ``non-cold'' transfer functions in Refs~\cite{Murgia:2017lwo,Murgia:2018now}.

Particularly significant in our results is that the 7 keV mass scale, potentially explaining the 3.5 keV line, maps onto WDM solutions to structure formation at the $\sim$2 keV thermal WDM particle mass \cite{Anderhalden:2012jc,Abazajian:2014gza} for this production scenario. The Lyman-alpha forest places constraints on thermal particle mass between $m_\mathrm{WDM} > 2.2\,\mathrm{keV}$ and $m_\mathrm{WDM} > 3.6\,\mathrm{keV}\ (2\sigma)$, depending on the freedom allowed in the thermal history of the intergalactic medium \cite{Garzilli:2015iwa,Murgia:2018now,Garzilli:2018jqh}. Galaxy counts may be a more robust measure of effects of WDM on small scale structure: current limits are at the $m_\mathrm{WDM}\gtrsim 2\, \mathrm{keV}$ scale. Such limits may become much more stringent, or bear evidence for reduced small-scale structure, as the Sloan Digital Sky Survey, the Dark Energy Survey and, in the future, the Large Synoptic Survey Telescope, will increase the reach of discovery of Local Group dwarf galaxies \cite{Cherry:2017dwu}.

\subsection{Production in the GUT Scale Scenario 1}
First order phase transition breaking $U(1)_{B-L}$ at a high scale needs to inject a higher amount of entropy in order to produce the required amount of dark matter, in the GUT scale scenario 1 described above.  This means that the momenta of the dark-matter sterile neutrinos are red-shifted by an additional factor, of 
$T_f/T_t=5$ due to the phase transition. This amount is in addition to the QCD-era dilution.  So, the free-streaming length is $\sim$5 times shorter than in the case of the scalar Higgs decay in the previous section. 
The dilution at the high-energy scale causes a redshifting by factor $\sim$5 in the momentum distribution of the sterile neutrino dark matter.

Due to the even colder nature of the GUT-scale sterile neutrino dark matter production, the cutoff scales for structure growth in this model are smaller than in the scalar decay model. The transfer functions for the GUT-scale production are best matched by equivalent thermal WDM masses of $1.1, 2.6, 4.9, 8.4 \rm\, keV$ for the $m_s = 1, 3, 7,\text{and}\ 14\,\mathrm{keV}$ cases, respectively, using the same methods as for the Higgs decay sterile neutrinos. 

\begin{figure}
    \centering
    \includegraphics[width=3.4in]{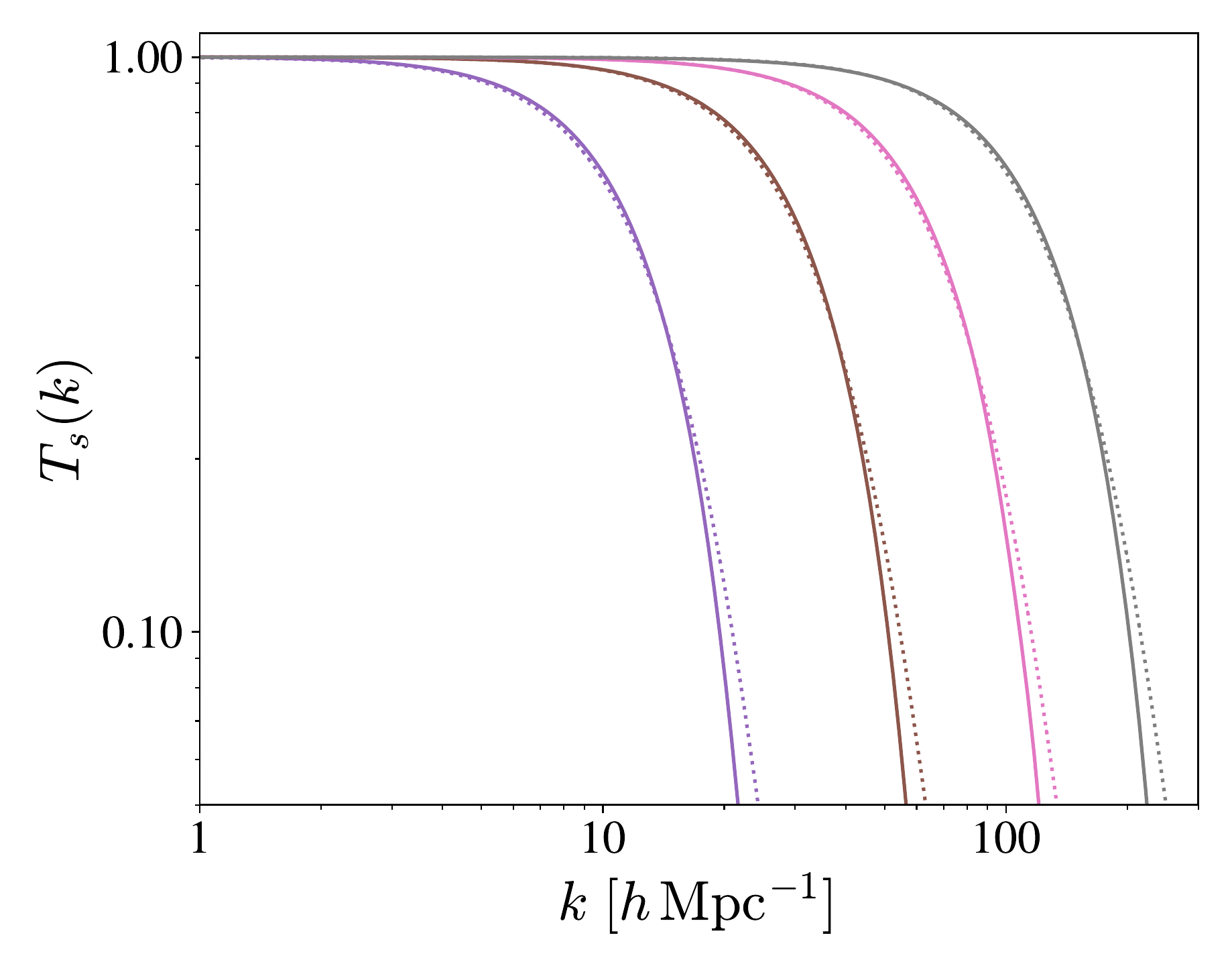}
    \caption{Transfer functions of sterile neutrino dark matter in the case of the sterile neutrino production at the GUT scale in scenario 1. The transfer functions $T_s(k)$ are shown for increasing cut-off wave number $k$ with increasing mass $m_s = 1, 3, 7,\text{and}\ 14\,\mathrm{keV}$. The equivalent thermal WDM particle mass equivalent transfer functions are shown in dashed lines, and correspond to $m_\mathrm{WDM} =  1.1, 2.6, 4.9, 8.4 \rm\, keV$, respectively.}
    \label{fig:KTY}
\end{figure}

\section{\boldmath Sterile neutrinos as a fraction of the dark matter: Multicomponent Dark Matter, X-ray Lines and the \texorpdfstring{$\sigma_8$}{sigma-8} Problem}
\label{FractionSection}

In scenarios of nonresonant and resonant production, where the full dark matter fraction is in sterile neutrinos, for a given mass (and lepton number, in the case of resonant production), the mixing angle is uniquely determined by requiring to reproduce the total dark matter density. However, if the fraction of dark matter as sterile neutrinos, $f\equiv \Omega_s/\Omega_\mathrm{DM}$, is smaller than one, the phenomenology of X-ray astronomy and effects on structure formation get more varied. In the case of nonresonant production, reducing $f$ proceeds from reducing the mixing angle at a given mass. In the case of resonant production, a reduction of $f$ comes from reducing the lepton number driving the resonance, or by lowering the mixing angle (in most regions of parameter space). In the case of scalar Higgs production, producing a fraction of dark matter breaks the miracle values of the production mechanism described above, but a small deviation from $f=1$ preserves the ``miracle.''   The GUT production scenarios are amenable to parameter variation to provide the full dark matter as sterile neutrinos to an arbitrary fractional dark matter case. The remainder of dark matter could be any cold component for the case of a cold plus warm dark matter (CWDM). Such a scenario may alleviate issues in galaxy formation by lowering the densities of substructure at the Local Group of galaxies scale \cite{Anderhalden:2012qt}. Significantly, a single sterile neutrino could act as both the cold and warm component, with the cold component produced via Higgs decay or GUT-scale production, and the warm component via nonresonant or resonant oscillations. 

\subsection{X-ray Lines}
It has been known for some time that sterile neutrino dark matter can produce an observable signature through the emission of X-ray lines \cite{Abazajian:2001nj,Abazajian:2001vt}. The observation of an unidentified X-ray line does not necessarily imply that sterile neutrinos make up all of the dark matter. In the case of nonresonant Dodelson-Widrow production, the energy of a line and its flux uniquely specifies the fraction of dark matter in sterile neutrinos required to produce the line. In the case of the 3.55 keV line seen toward stacked clusters \cite{Bulbul:2014sua}, the particle mass and mixing angle required to produce the line energy and flux specifies a dark matter density that is $\sim 13\%$ of the full dark matter density, via the production relation, Eq.~7 in Ref.~\cite{Abazajian:2005gj}. 

In the case of resonant production, the lepton number can be reduced, decreasing resonant production so that the mixing angle required is larger to give the correct density, but also reducing the fraction of dark matter in sterile neutrinos. This produces a range of mixing angles from $\sin^2 2\theta \approx 7\times 10^{-11}$ to $\sin^2 2\theta \approx 5\times 10^{-10}$,  while the lepton number ranges  from $7\times 10^{-5}$ to zero, as the mixing angle goes from the 100\% resonant production at the smallest mixing angles to that of the zero lepton number (Dodelson-Widrow) case at $\sin^2 2\theta \approx 5\times 10^{-10}$. If we consider a low-reheating temperature scenario \cite{Gelmini:2004ah}, scattering production is reduced and the mixing angle could even be larger, up to $\sin^2 2\theta \approx 10^{-7}$ so that the observed 3.55 keV line can be explained with a fraction $f \approx 7\times 10^{-4}$ \cite{Abazajian:2017tcc}.

Two important observations can be made with respect to X-ray signals from sterile neutrinos that contribute fraction $f<1$ of the total dark matter. 
First, the mixing angle must be a factor $1/f$ larger, which for the case of the 3.55 keV line, is a factor $\sim$7 larger than for the case $f=1$. This means the mixing angle can be as large as  $\sin^2 2\theta \approx 5\times 10^{-10}$ in the case of the stacked cluster {\it XMM-Newton} MOS instrument detection \cite{Bulbul:2014sua}. Second, a fractional sterile neutrino dark matter scenario is not subject to the same small-scale structure constraints.  If $f<1$, and the rest of dark matter is cold, the constraints derived for pure WDM can be  alleviated since the small scale structure clustering is preserved by the predominance of cold dark matter~\cite{Boyanovsky:2007ba,Petraki:2008ef,Boyarsky:2008xj,Anderhalden:2012jc}. 

\subsection{\boldmath The \texorpdfstring{$\sigma_8$}{sigma-8} Problem}
At a different scale from where sterile neutrinos can be all of the dark matter, $m_s \lesssim 1\ \mathrm{keV}$ a fraction of sterile neutrinos to dark matter may resolve a persistent cosmological tension. That tension is between the cosmic microwave background (CMB) inferences and local Universe measures of the amplitude of the matter power spectrum at $\sim 8\,{\rm Mpc}/h$, dubbed $\sigma_8$ \cite{Wyman:2013lza,Battye:2013xqa,Dvorkin:2014lea,Beutler:2014yhv,Canac:2016smv}. The constraints are often degenerate with the density of matter, $\Omega_m$, quantified in terms of some combination as $S_8 = \sigma_8(\Omega_m/0.3)^x$ for some power $x$. Joint fits between the early and late Universe require the value of $\Omega_m$ to be identical given a unified model, but a modification of the power spectrum amplitude between large and small scales is needed in order to reconcile $S_8$ (or, equivalently, $\sigma_8$). 

We point out here that the fractional production of sterile neutrino dark matter, via any of the above mechanisms, could reduce the amplitude of the power spectrum at the appropriate scale larger than or near $8\,\mathrm{Mpc}/h$, reducing $\sigma_8$ without altering large scales constrained by the CMB. The suppression of the power spectrum by a warm fraction of dark matter as WDM, while the remainder is cold dark matter (CDM), in a cold-plus-warm dark matter model (CWDM) was calculated in Ref.~\cite{Boyarsky:2008xj}. They found it to be a plateau of 
\begin{equation}
    T_\mathrm{plateau} = \sqrt{\frac{P_\mathrm{CWDM}}{P_\mathrm{CDM}}} \approx 1-14 f_\mathrm{WDM}, \quad \text{for\ } k\gg k_{fs}
    \label{eq:fwdm}
\end{equation}
where $f_\mathrm{WDM}\equiv \Omega_\mathrm{WDM}/\Omega_m$, and $\Omega_m = \Omega_\mathrm{CDM}+\Omega_\mathrm{WDM}+\Omega_\mathrm{baryon}$ (and different from $f$ defined above). Note that this expression, Eq.~\eqref{eq:fwdm}, is an expansion which only applies when $f_\mathrm{WDM}$ is small. The WDM free streaming scale in wavenumber is defined as \cite{Boyarsky:2008xj}
\begin{equation}
k_{fs} \equiv \sqrt{4\pi G\rho}\frac{a}{\langle v\rangle} = \sqrt\frac{3}{2}\frac{aH}{\langle v\rangle},
\end{equation}
where $\rho$ is the total density of the universe through the time considered, $\langle v\rangle$ is the velocity dispersion of the WDM component only, $a$ is the scale factor, and $H$ is the Hubble rate.
We verified the relation Eq.~\eqref{eq:fwdm} with a modified version of CAMB \cite{Lewis:1999bs}. 

In order to match the amplitude and shape of the matter power spectrum inferred from the CMB, measured at larger scales, the value of $\sigma_8$ should be reduced by 5\% to 15\%, depending on the combination of data sets at smaller scales. This requires two conditions: a fraction of dark matter as WDM of approximately $f_\mathrm{WDM} \approx 0.4\%\ \text{to}\ 1\%$, and a free streaming length of the dark matter larger than, or of order, 8 Mpc$/h$  but below scales affecting the primary CMB, which corresponds to an effective thermal WDM particle mass of $40\,\mathrm{eV}\lesssim m_\mathrm{thermal} \lesssim 60\,\mathrm{eV}$. This is a relatively narrow prediction range for the new dark matter particle. 

Since the particle production method only affects the shape of the transfer function near $\lambda_{fs}$ (or $k_{fs}$), and the plateau of suppression is simply that which remains , the mechanism that produces 0.4\% to 1\% of the dark matter as WDM is not crucial in resolving the $\sigma_8$ problem. However, as an example, the Dodelson-Widrow mechanism would predict a sterile neutrino particle mass of 60 eV to 100 eV, and a mixing angle of approximately $\sin^2 2\theta \approx 8\times 10^{-8}$ to $2\times 10^{-7}$ \cite{Abazajian:2001nj,Abazajian:2005gj}, which could be within the sensitivity of the tritium beta-decay experiment KATRIN \cite{Mertens:2014nha}. For the resonant Shi-Fuller mechanism, the mixing angles of $\sin^2 2\theta \sim 10^{-12}$ to $\sim 10^{-7}$ would produce the proper density of dark matter at sterile neutrino  mass of 60 eV to 100 eV (cf.\ Fig.\ 2 in Ref.~\cite{Abazajian:2001nj}). For production via singlet Higgs decay or GUT high-temperature production, the mixing angle is bounded from above by the Dodelson-Widrow values, since production by oscillations is required, yet the mixing angle(s) could be smaller. The requisite properties of the necessary sterile neutrinos described here would produce a line in the extreme ultraviolet through their radiative decay, at $E_\gamma \approx 30-50\ \mathrm{eV}$ or $\lambda \approx 25-40\ \mathrm{nm}$. Unfortunately, this window remains largely unexplored, below the wavelength range of Hubble Space Telescope UV instruments as well as the Galaxy Evolution Explorer (GALEX), though this wavelength is within the observed range of the decommissioned Extreme Ultraviolet Explorer (EUVE) \cite{1995ApJ...442..653J}.

\section{Conclusions}

In this paper, we have analyzed the nature of a wide range  of sterile neutrino dark matter production scenarios. The production mechanism determines a range of structure formation signatures and their relation with other properties of the sterile neutrinos. In addition, the possibility that a fraction $f<1$ of dark matter is in sterile neutrinos opens up a larger range of possibilities.  In this case, structure formation can have the features of a broader cold plus warm dark matter models, and the warm component alleviates the so-called $\sigma_8$ problem. 

Let us summarize our most significant conclusions.
First, a varied set of mechanisms exist to produce dark matter. The Higgs scalar decay model produces a ``miracle'' density for the standard choice of parameters, while other models require a tuning of a parameter to match the observed or fractionally inferred dark matter abundance. Second, the production models yield a wide range of WDM cutoff scales that range from exceedingly large to beyond current constraints on WDM.  Third, constraints on the free-streaming scale in models that produce the 3.55 keV line could indicate a high-temperature GUT scale scenario, where the free streaming scale is equivalent to that of a 4.9 keV thermal WDM particle, at or beyond the limit of the strongest claimed structure formation constraints, such as the high-resolution Lyman-$\alpha$ forest \cite{Irsic:2017ixq}. Fourth, free streaming scale constraints could indicate that only a fraction of the dark matter may be consitituted of warm sterile neutrinos, which partially suppress structure formation, that however continues to be driven mainly by CDM or another strongly clustering variant. This would evade pure WDM structure formation limits. Fifth, an X-ray line does not require any production mechanism to be responsible for all of the observed dark matter, and fractions of as little as $\sim 10^{-3}$ to $\sim$13\% could explain the 3.5 keV line. And, last but not least,   the case of fractional production could work with small particle masses, corresponding to free-streaming  at appropriately large scales and abundances that match the reduction of power at small scales, consequently alleviating the $\sigma_8$ problem.   

Overall, a sterile neutrino related to the mass generation mechanism for the active neutrinos remains an intriguing candidate for dark matter. 
The effects of such a sterile neutrino constituting all or part of the dark matter influences X-ray astronomy, cosmological and galactic structure formation, as well as nuclear physics.
Sterile neutrino dark matter properties can be inferred directly or indirectly through
multiple methods. These methods may be the key to unveil its existence, its ties to the high energy mechanism from which the sterile neutrino is originated, and the mechanism which drives its production in the early universe.

\begin{acknowledgments}
K.N.A. was supported by NSF Theoretical Physics Grant No. PHY-1620638. A.K. was supported by the U.S. Department of Energy Grant No. DE-SC0009937 and by the World Premier International Research Center Initiative (WPI), MEXT, Japan. We acknowledge the Simons Foundation's 2018 Simons Symposium on ``Illuminating Dark Matter'' where initial discussions of this work started, and an anonymous referee for detailed comments.
\end{acknowledgments}

\bibliography{sterile}

\begin{thebibliography}{43}%
\makeatletter
\providecommand \@ifxundefined [1]{%
 \@ifx{#1\undefined}
}%
\providecommand \@ifnum [1]{%
 \ifnum #1\expandafter \@firstoftwo
 \else \expandafter \@secondoftwo
 \fi
}%
\providecommand \@ifx [1]{%
 \ifx #1\expandafter \@firstoftwo
 \else \expandafter \@secondoftwo
 \fi
}%
\providecommand \natexlab [1]{#1}%
\providecommand \enquote  [1]{``#1''}%
\providecommand \bibnamefont  [1]{#1}%
\providecommand \bibfnamefont [1]{#1}%
\providecommand \citenamefont [1]{#1}%
\providecommand \href@noop [0]{\@secondoftwo}%
\providecommand \href [0]{\begingroup \@sanitize@url \@href}%
\providecommand \@href[1]{\@@startlink{#1}\@@href}%
\providecommand \@@href[1]{\endgroup#1\@@endlink}%
\providecommand \@sanitize@url [0]{\catcode `\\12\catcode `\$12\catcode
  `\&12\catcode `\#12\catcode `\^12\catcode `\_12\catcode `\%12\relax}%
\providecommand \@@startlink[1]{}%
\providecommand \@@endlink[0]{}%
\providecommand \url  [0]{\begingroup\@sanitize@url \@url }%
\providecommand \@url [1]{\endgroup\@href {#1}{\urlprefix }}%
\providecommand \urlprefix  [0]{URL }%
\providecommand \Eprint [0]{\href }%
\providecommand \doibase [0]{http://dx.doi.org/}%
\providecommand \selectlanguage [0]{\@gobble}%
\providecommand \bibinfo  [0]{\@secondoftwo}%
\providecommand \bibfield  [0]{\@secondoftwo}%
\providecommand \translation [1]{[#1]}%
\providecommand \BibitemOpen [0]{}%
\providecommand \bibitemStop [0]{}%
\providecommand \bibitemNoStop [0]{.\EOS\space}%
\providecommand \EOS [0]{\spacefactor3000\relax}%
\providecommand \BibitemShut  [1]{\csname bibitem#1\endcsname}%
\let\auto@bib@innerbib\@empty
\bibitem [{\citenamefont {Kusenko}(2009)}]{Kusenko:2009up}%
  \BibitemOpen
  \bibfield  {author} {\bibinfo {author} {\bibfnamefont {A.}~\bibnamefont
  {Kusenko}},\ }\href {\doibase 10.1016/j.physrep.2009.07.004} {\bibfield
  {journal} {\bibinfo  {journal} {Phys. Rept.}\ }\textbf {\bibinfo {volume}
  {481}},\ \bibinfo {pages} {1} (\bibinfo {year} {2009})},\ \Eprint
  {http://arxiv.org/abs/0906.2968} {arXiv:0906.2968 [hep-ph]} \BibitemShut
  {NoStop}%
\bibitem [{\citenamefont {Dodelson}\ and\ \citenamefont
  {Widrow}(1994)}]{Dodelson:1993je}%
  \BibitemOpen
  \bibfield  {author} {\bibinfo {author} {\bibfnamefont {S.}~\bibnamefont
  {Dodelson}}\ and\ \bibinfo {author} {\bibfnamefont {L.~M.}\ \bibnamefont
  {Widrow}},\ }\href@noop {} {\bibfield  {journal} {\bibinfo  {journal} {Phys.
  Rev. Lett.}\ }\textbf {\bibinfo {volume} {72}},\ \bibinfo {pages} {17}
  (\bibinfo {year} {1994})},\ \Eprint {http://arxiv.org/abs/hep-ph/9303287}
  {hep-ph/9303287} \BibitemShut {NoStop}%
\bibitem [{\citenamefont {Abazajian}\ \emph
  {et~al.}(2001{\natexlab{a}})\citenamefont {Abazajian}, \citenamefont
  {Fuller},\ and\ \citenamefont {Patel}}]{Abazajian:2001nj}%
  \BibitemOpen
  \bibfield  {author} {\bibinfo {author} {\bibfnamefont {K.}~\bibnamefont
  {Abazajian}}, \bibinfo {author} {\bibfnamefont {G.~M.}\ \bibnamefont
  {Fuller}}, \ and\ \bibinfo {author} {\bibfnamefont {M.}~\bibnamefont
  {Patel}},\ }\href@noop {} {\bibfield  {journal} {\bibinfo  {journal} {Phys.
  Rev.}\ }\textbf {\bibinfo {volume} {D64}},\ \bibinfo {pages} {023501}
  (\bibinfo {year} {2001}{\natexlab{a}})},\ \Eprint
  {http://arxiv.org/abs/astro-ph/0101524} {astro-ph/0101524} \BibitemShut
  {NoStop}%
\bibitem [{\citenamefont {Kusenko}\ and\ \citenamefont
  {Segr{\`e}}(1997)}]{Kusenko:1997sp}%
  \BibitemOpen
  \bibfield  {author} {\bibinfo {author} {\bibfnamefont {A.}~\bibnamefont
  {Kusenko}}\ and\ \bibinfo {author} {\bibfnamefont {G.}~\bibnamefont
  {Segr{\`e}}},\ }\href {\doibase 10.1016/S0370-2693(97)00121-4} {\bibfield
  {journal} {\bibinfo  {journal} {Phys. Lett.}\ }\textbf {\bibinfo {volume}
  {B396}},\ \bibinfo {pages} {197} (\bibinfo {year} {1997})},\ \Eprint
  {http://arxiv.org/abs/hep-ph/9701311} {arXiv:hep-ph/9701311} \BibitemShut
  {NoStop}%
\bibitem [{\citenamefont {Fuller}\ \emph {et~al.}(2003)\citenamefont {Fuller},
  \citenamefont {Kusenko}, \citenamefont {Mocioiu},\ and\ \citenamefont
  {Pascoli}}]{Fuller:2003gy}%
  \BibitemOpen
  \bibfield  {author} {\bibinfo {author} {\bibfnamefont {G.~M.}\ \bibnamefont
  {Fuller}}, \bibinfo {author} {\bibfnamefont {A.}~\bibnamefont {Kusenko}},
  \bibinfo {author} {\bibfnamefont {I.}~\bibnamefont {Mocioiu}}, \ and\
  \bibinfo {author} {\bibfnamefont {S.}~\bibnamefont {Pascoli}},\ }\href@noop
  {} {\bibfield  {journal} {\bibinfo  {journal} {Phys. Rev.}\ }\textbf
  {\bibinfo {volume} {D68}},\ \bibinfo {pages} {103002} (\bibinfo {year}
  {2003})},\ \Eprint {http://arxiv.org/abs/astro-ph/0307267} {astro-ph/0307267}
  \BibitemShut {NoStop}%
\bibitem [{\citenamefont {Asaka}\ \emph {et~al.}(2005)\citenamefont {Asaka},
  \citenamefont {Blanchet},\ and\ \citenamefont {Shaposhnikov}}]{Asaka:2005an}%
  \BibitemOpen
  \bibfield  {author} {\bibinfo {author} {\bibfnamefont {T.}~\bibnamefont
  {Asaka}}, \bibinfo {author} {\bibfnamefont {S.}~\bibnamefont {Blanchet}}, \
  and\ \bibinfo {author} {\bibfnamefont {M.}~\bibnamefont {Shaposhnikov}},\
  }\href@noop {} {\bibfield  {journal} {\bibinfo  {journal} {Phys. Lett.}\
  }\textbf {\bibinfo {volume} {B631}},\ \bibinfo {pages} {151} (\bibinfo {year}
  {2005})},\ \Eprint {http://arxiv.org/abs/hep-ph/0503065} {hep-ph/0503065}
  \BibitemShut {NoStop}%
\bibitem [{\citenamefont {Asaka}\ and\ \citenamefont
  {Shaposhnikov}(2005)}]{Asaka:2005pn}%
  \BibitemOpen
  \bibfield  {author} {\bibinfo {author} {\bibfnamefont {T.}~\bibnamefont
  {Asaka}}\ and\ \bibinfo {author} {\bibfnamefont {M.}~\bibnamefont
  {Shaposhnikov}},\ }\href@noop {} {\bibfield  {journal} {\bibinfo  {journal}
  {Phys. Lett.}\ }\textbf {\bibinfo {volume} {B620}},\ \bibinfo {pages} {17}
  (\bibinfo {year} {2005})},\ \Eprint {http://arxiv.org/abs/hep-ph/0505013}
  {hep-ph/0505013} \BibitemShut {NoStop}%
\bibitem [{\citenamefont {Abazajian}\ \emph
  {et~al.}(2001{\natexlab{b}})\citenamefont {Abazajian}, \citenamefont
  {Fuller},\ and\ \citenamefont {Tucker}}]{Abazajian:2001vt}%
  \BibitemOpen
  \bibfield  {author} {\bibinfo {author} {\bibfnamefont {K.}~\bibnamefont
  {Abazajian}}, \bibinfo {author} {\bibfnamefont {G.~M.}\ \bibnamefont
  {Fuller}}, \ and\ \bibinfo {author} {\bibfnamefont {W.~H.}\ \bibnamefont
  {Tucker}},\ }\href@noop {} {\bibfield  {journal} {\bibinfo  {journal}
  {Astrophys. J.}\ }\textbf {\bibinfo {volume} {562}},\ \bibinfo {pages} {593}
  (\bibinfo {year} {2001}{\natexlab{b}})},\ \Eprint
  {http://arxiv.org/abs/astro-ph/0106002} {astro-ph/0106002} \BibitemShut
  {NoStop}%
\bibitem [{\citenamefont {Kusenko}(2006)}]{Kusenko:2006rh}%
  \BibitemOpen
  \bibfield  {author} {\bibinfo {author} {\bibfnamefont {A.}~\bibnamefont
  {Kusenko}},\ }\href@noop {} {\bibfield  {journal} {\bibinfo  {journal} {Phys.
  Rev. Lett.}\ }\textbf {\bibinfo {volume} {97}},\ \bibinfo {pages} {241301}
  (\bibinfo {year} {2006})},\ \Eprint {http://arxiv.org/abs/hep-ph/0609081}
  {hep-ph/0609081} \BibitemShut {NoStop}%
\bibitem [{\citenamefont {Petraki}\ and\ \citenamefont
  {Kusenko}(2008)}]{Petraki:2007gq}%
  \BibitemOpen
  \bibfield  {author} {\bibinfo {author} {\bibfnamefont {K.}~\bibnamefont
  {Petraki}}\ and\ \bibinfo {author} {\bibfnamefont {A.}~\bibnamefont
  {Kusenko}},\ }\href {\doibase 10.1103/PhysRevD.77.065014} {\bibfield
  {journal} {\bibinfo  {journal} {Phys. Rev.}\ }\textbf {\bibinfo {volume}
  {D77}},\ \bibinfo {pages} {065014} (\bibinfo {year} {2008})},\ \Eprint
  {http://arxiv.org/abs/arXiv:0711.4646} {arXiv:arXiv:0711.4646 [hep-ph]}
  \BibitemShut {NoStop}%
\bibitem [{\citenamefont {Kang}(2015)}]{Kang:2014cia}%
  \BibitemOpen
  \bibfield  {author} {\bibinfo {author} {\bibfnamefont {Z.}~\bibnamefont
  {Kang}},\ }\href {\doibase 10.1140/epjc/s10052-015-3702-4} {\bibfield
  {journal} {\bibinfo  {journal} {Eur. Phys. J.}\ }\textbf {\bibinfo {volume}
  {C75}},\ \bibinfo {pages} {471} (\bibinfo {year} {2015})},\ \Eprint
  {http://arxiv.org/abs/1411.2773} {arXiv:1411.2773 [hep-ph]} \BibitemShut
  {NoStop}%
\bibitem [{\citenamefont {Shaposhnikov}\ and\ \citenamefont
  {Tkachev}(2006)}]{Shaposhnikov:2006xi}%
  \BibitemOpen
  \bibfield  {author} {\bibinfo {author} {\bibfnamefont {M.}~\bibnamefont
  {Shaposhnikov}}\ and\ \bibinfo {author} {\bibfnamefont {I.}~\bibnamefont
  {Tkachev}},\ }\href@noop {} {\bibfield  {journal} {\bibinfo  {journal} {Phys.
  Lett.}\ }\textbf {\bibinfo {volume} {B639}},\ \bibinfo {pages} {414}
  (\bibinfo {year} {2006})},\ \Eprint {http://arxiv.org/abs/hep-ph/0604236}
  {hep-ph/0604236} \BibitemShut {NoStop}%
\bibitem [{\citenamefont {Kusenko}\ \emph {et~al.}(2010)\citenamefont
  {Kusenko}, \citenamefont {Takahashi},\ and\ \citenamefont
  {Yanagida}}]{Kusenko:2010ik}%
  \BibitemOpen
  \bibfield  {author} {\bibinfo {author} {\bibfnamefont {A.}~\bibnamefont
  {Kusenko}}, \bibinfo {author} {\bibfnamefont {F.}~\bibnamefont {Takahashi}},
  \ and\ \bibinfo {author} {\bibfnamefont {T.~T.}\ \bibnamefont {Yanagida}},\
  }\href {\doibase 10.1016/j.physletb.2010.08.031} {\bibfield  {journal}
  {\bibinfo  {journal} {Phys. Lett.}\ }\textbf {\bibinfo {volume} {B693}},\
  \bibinfo {pages} {144} (\bibinfo {year} {2010})},\ \Eprint
  {http://arxiv.org/abs/1006.1731} {arXiv:1006.1731 [hep-ph]} \BibitemShut
  {NoStop}%
\bibitem [{\citenamefont {Abazajian}(2006)}]{Abazajian:2005gj}%
  \BibitemOpen
  \bibfield  {author} {\bibinfo {author} {\bibfnamefont {K.}~\bibnamefont
  {Abazajian}},\ }\href@noop {} {\bibfield  {journal} {\bibinfo  {journal}
  {Phys. Rev.}\ }\textbf {\bibinfo {volume} {D73}},\ \bibinfo {pages} {063506}
  (\bibinfo {year} {2006})},\ \Eprint {http://arxiv.org/abs/astro-ph/0511630}
  {astro-ph/0511630} \BibitemShut {NoStop}%
\bibitem [{\citenamefont {Horiuchi}\ \emph {et~al.}(2014)\citenamefont
  {Horiuchi}, \citenamefont {Humphrey}, \citenamefont {Onorbe}, \citenamefont
  {Abazajian}, \citenamefont {Kaplinghat},\ and\ \citenamefont
  {Garrison-Kimmel}}]{Horiuchi:2013noa}%
  \BibitemOpen
  \bibfield  {author} {\bibinfo {author} {\bibfnamefont {S.}~\bibnamefont
  {Horiuchi}}, \bibinfo {author} {\bibfnamefont {P.~J.}\ \bibnamefont
  {Humphrey}}, \bibinfo {author} {\bibfnamefont {J.}~\bibnamefont {Onorbe}},
  \bibinfo {author} {\bibfnamefont {K.~N.}\ \bibnamefont {Abazajian}}, \bibinfo
  {author} {\bibfnamefont {M.}~\bibnamefont {Kaplinghat}}, \ and\ \bibinfo
  {author} {\bibfnamefont {S.}~\bibnamefont {Garrison-Kimmel}},\ }\href
  {\doibase 10.1103/PhysRevD.89.025017} {\bibfield  {journal} {\bibinfo
  {journal} {Phys. Rev.}\ }\textbf {\bibinfo {volume} {D89}},\ \bibinfo {pages}
  {025017} (\bibinfo {year} {2014})},\ \Eprint {http://arxiv.org/abs/1311.0282}
  {arXiv:1311.0282 [astro-ph.CO]} \BibitemShut {NoStop}%
\bibitem [{\citenamefont {Bulbul}\ \emph {et~al.}(2014)\citenamefont {Bulbul},
  \citenamefont {Markevitch}, \citenamefont {Foster}, \citenamefont {Smith},
  \citenamefont {Loewenstein},\ and\ \citenamefont {Randall}}]{Bulbul:2014sua}%
  \BibitemOpen
  \bibfield  {author} {\bibinfo {author} {\bibfnamefont {E.}~\bibnamefont
  {Bulbul}}, \bibinfo {author} {\bibfnamefont {M.}~\bibnamefont {Markevitch}},
  \bibinfo {author} {\bibfnamefont {A.}~\bibnamefont {Foster}}, \bibinfo
  {author} {\bibfnamefont {R.~K.}\ \bibnamefont {Smith}}, \bibinfo {author}
  {\bibfnamefont {M.}~\bibnamefont {Loewenstein}}, \ and\ \bibinfo {author}
  {\bibfnamefont {S.~W.}\ \bibnamefont {Randall}},\ }\href {\doibase
  10.1088/0004-637X/789/1/13} {\bibfield  {journal} {\bibinfo  {journal}
  {Astrophys. J.}\ }\textbf {\bibinfo {volume} {789}},\ \bibinfo {pages} {13}
  (\bibinfo {year} {2014})},\ \Eprint {http://arxiv.org/abs/1402.2301}
  {arXiv:1402.2301 [astro-ph.CO]} \BibitemShut {NoStop}%
\bibitem [{\citenamefont {Boyarsky}\ \emph {et~al.}(2014)\citenamefont
  {Boyarsky}, \citenamefont {Ruchayskiy}, \citenamefont {Iakubovskyi},\ and\
  \citenamefont {Franse}}]{Boyarsky:2014jta}%
  \BibitemOpen
  \bibfield  {author} {\bibinfo {author} {\bibfnamefont {A.}~\bibnamefont
  {Boyarsky}}, \bibinfo {author} {\bibfnamefont {O.}~\bibnamefont
  {Ruchayskiy}}, \bibinfo {author} {\bibfnamefont {D.}~\bibnamefont
  {Iakubovskyi}}, \ and\ \bibinfo {author} {\bibfnamefont {J.}~\bibnamefont
  {Franse}},\ }\href {\doibase 10.1103/PhysRevLett.113.251301} {\bibfield
  {journal} {\bibinfo  {journal} {Phys. Rev. Lett.}\ }\textbf {\bibinfo
  {volume} {113}},\ \bibinfo {pages} {251301} (\bibinfo {year} {2014})},\
  \Eprint {http://arxiv.org/abs/1402.4119} {arXiv:1402.4119 [astro-ph.CO]}
  \BibitemShut {NoStop}%
\bibitem [{\citenamefont {Diamanti}\ \emph {et~al.}(2017)\citenamefont
  {Diamanti}, \citenamefont {Ando}, \citenamefont {Gariazzo}, \citenamefont
  {Mena},\ and\ \citenamefont {Weniger}}]{Diamanti:2017xfo}%
  \BibitemOpen
  \bibfield  {author} {\bibinfo {author} {\bibfnamefont {R.}~\bibnamefont
  {Diamanti}}, \bibinfo {author} {\bibfnamefont {S.}~\bibnamefont {Ando}},
  \bibinfo {author} {\bibfnamefont {S.}~\bibnamefont {Gariazzo}}, \bibinfo
  {author} {\bibfnamefont {O.}~\bibnamefont {Mena}}, \ and\ \bibinfo {author}
  {\bibfnamefont {C.}~\bibnamefont {Weniger}},\ }\href {\doibase
  10.1088/1475-7516/2017/06/008} {\bibfield  {journal} {\bibinfo  {journal}
  {JCAP}\ }\textbf {\bibinfo {volume} {1706}},\ \bibinfo {pages} {008}
  (\bibinfo {year} {2017})},\ \Eprint {http://arxiv.org/abs/1701.03128}
  {arXiv:1701.03128 [astro-ph.CO]} \BibitemShut {NoStop}%
\bibitem [{\citenamefont {Abazajian}(2017)}]{Abazajian:2017tcc}%
  \BibitemOpen
  \bibfield  {author} {\bibinfo {author} {\bibfnamefont {K.~N.}\ \bibnamefont
  {Abazajian}},\ }\href {\doibase 10.1016/j.physrep.2017.10.003} {\bibfield
  {journal} {\bibinfo  {journal} {Phys. Rept.}\ }\textbf {\bibinfo {volume}
  {711-712}},\ \bibinfo {pages} {1} (\bibinfo {year} {2017})},\ \Eprint
  {http://arxiv.org/abs/1705.01837} {arXiv:1705.01837 [hep-ph]} \BibitemShut
  {NoStop}%
\bibitem [{\citenamefont {Shi}\ and\ \citenamefont
  {Fuller}(1999)}]{Shi:1998km}%
  \BibitemOpen
  \bibfield  {author} {\bibinfo {author} {\bibfnamefont {X.-D.}\ \bibnamefont
  {Shi}}\ and\ \bibinfo {author} {\bibfnamefont {G.~M.}\ \bibnamefont
  {Fuller}},\ }\href@noop {} {\bibfield  {journal} {\bibinfo  {journal} {Phys.
  Rev. Lett.}\ }\textbf {\bibinfo {volume} {82}},\ \bibinfo {pages} {2832}
  (\bibinfo {year} {1999})},\ \Eprint {http://arxiv.org/abs/astro-ph/9810076}
  {astro-ph/9810076} \BibitemShut {NoStop}%
\bibitem [{\citenamefont {Venumadhav}\ \emph {et~al.}(2016)\citenamefont
  {Venumadhav}, \citenamefont {Cyr-Racine}, \citenamefont {Abazajian},\ and\
  \citenamefont {Hirata}}]{Venumadhav:2015pla}%
  \BibitemOpen
  \bibfield  {author} {\bibinfo {author} {\bibfnamefont {T.}~\bibnamefont
  {Venumadhav}}, \bibinfo {author} {\bibfnamefont {F.-Y.}\ \bibnamefont
  {Cyr-Racine}}, \bibinfo {author} {\bibfnamefont {K.~N.}\ \bibnamefont
  {Abazajian}}, \ and\ \bibinfo {author} {\bibfnamefont {C.~M.}\ \bibnamefont
  {Hirata}},\ }\href {\doibase 10.1103/PhysRevD.94.043515} {\bibfield
  {journal} {\bibinfo  {journal} {Phys. Rev.}\ }\textbf {\bibinfo {volume}
  {D94}},\ \bibinfo {pages} {043515} (\bibinfo {year} {2016})},\ \Eprint
  {http://arxiv.org/abs/1507.06655} {arXiv:1507.06655 [astro-ph.CO]}
  \BibitemShut {NoStop}%
\bibitem [{\citenamefont {Abazajian}(2014)}]{Abazajian:2014gza}%
  \BibitemOpen
  \bibfield  {author} {\bibinfo {author} {\bibfnamefont {K.~N.}\ \bibnamefont
  {Abazajian}},\ }\href {\doibase 10.1103/PhysRevLett.112.161303} {\bibfield
  {journal} {\bibinfo  {journal} {Phys. Rev. Lett.}\ }\textbf {\bibinfo
  {volume} {112}},\ \bibinfo {pages} {161303} (\bibinfo {year} {2014})},\
  \Eprint {http://arxiv.org/abs/1403.0954} {arXiv:1403.0954 [astro-ph.CO]}
  \BibitemShut {NoStop}%
\bibitem [{\citenamefont {Cherry}\ and\ \citenamefont
  {Horiuchi}(2017)}]{Cherry:2017dwu}%
  \BibitemOpen
  \bibfield  {author} {\bibinfo {author} {\bibfnamefont {J.~F.}\ \bibnamefont
  {Cherry}}\ and\ \bibinfo {author} {\bibfnamefont {S.}~\bibnamefont
  {Horiuchi}},\ }\href {\doibase 10.1103/PhysRevD.95.083015} {\bibfield
  {journal} {\bibinfo  {journal} {Phys. Rev.}\ }\textbf {\bibinfo {volume}
  {D95}},\ \bibinfo {pages} {083015} (\bibinfo {year} {2017})},\ \Eprint
  {http://arxiv.org/abs/1701.07874} {arXiv:1701.07874 [hep-ph]} \BibitemShut
  {NoStop}%
\bibitem [{\citenamefont {Lewis}\ \emph {et~al.}(2000)\citenamefont {Lewis},
  \citenamefont {Challinor},\ and\ \citenamefont {Lasenby}}]{Lewis:1999bs}%
  \BibitemOpen
  \bibfield  {author} {\bibinfo {author} {\bibfnamefont {A.}~\bibnamefont
  {Lewis}}, \bibinfo {author} {\bibfnamefont {A.}~\bibnamefont {Challinor}}, \
  and\ \bibinfo {author} {\bibfnamefont {A.}~\bibnamefont {Lasenby}},\ }\href
  {\doibase 10.1086/309179} {\bibfield  {journal} {\bibinfo  {journal} {\apj}\
  }\textbf {\bibinfo {volume} {538}},\ \bibinfo {pages} {473} (\bibinfo {year}
  {2000})},\ \Eprint {http://arxiv.org/abs/astro-ph/9911177}
  {arXiv:astro-ph/9911177 [astro-ph]} \BibitemShut {NoStop}%
\bibitem [{\citenamefont {Bode}\ \emph {et~al.}(2001)\citenamefont {Bode},
  \citenamefont {Ostriker},\ and\ \citenamefont {Turok}}]{Bode:2000gq}%
  \BibitemOpen
  \bibfield  {author} {\bibinfo {author} {\bibfnamefont {P.}~\bibnamefont
  {Bode}}, \bibinfo {author} {\bibfnamefont {J.~P.}\ \bibnamefont {Ostriker}},
  \ and\ \bibinfo {author} {\bibfnamefont {N.}~\bibnamefont {Turok}},\
  }\href@noop {} {\bibfield  {journal} {\bibinfo  {journal} {Astrophys. J.}\
  }\textbf {\bibinfo {volume} {556}},\ \bibinfo {pages} {93} (\bibinfo {year}
  {2001})},\ \Eprint {http://arxiv.org/abs/astro-ph/0010389} {astro-ph/0010389}
  \BibitemShut {NoStop}%
\bibitem [{\citenamefont {Murgia}\ \emph {et~al.}(2017)\citenamefont {Murgia},
  \citenamefont {Merle}, \citenamefont {Viel}, \citenamefont {Totzauer},\ and\
  \citenamefont {Schneider}}]{Murgia:2017lwo}%
  \BibitemOpen
  \bibfield  {author} {\bibinfo {author} {\bibfnamefont {R.}~\bibnamefont
  {Murgia}}, \bibinfo {author} {\bibfnamefont {A.}~\bibnamefont {Merle}},
  \bibinfo {author} {\bibfnamefont {M.}~\bibnamefont {Viel}}, \bibinfo {author}
  {\bibfnamefont {M.}~\bibnamefont {Totzauer}}, \ and\ \bibinfo {author}
  {\bibfnamefont {A.}~\bibnamefont {Schneider}},\ }\href {\doibase
  10.1088/1475-7516/2017/11/046} {\bibfield  {journal} {\bibinfo  {journal}
  {JCAP}\ }\textbf {\bibinfo {volume} {1711}},\ \bibinfo {pages} {046}
  (\bibinfo {year} {2017})},\ \Eprint {http://arxiv.org/abs/1704.07838}
  {arXiv:1704.07838 [astro-ph.CO]} \BibitemShut {NoStop}%
\bibitem [{\citenamefont {Murgia}\ \emph {et~al.}(2018)\citenamefont {Murgia},
  \citenamefont {Iršič},\ and\ \citenamefont {Viel}}]{Murgia:2018now}%
  \BibitemOpen
  \bibfield  {author} {\bibinfo {author} {\bibfnamefont {R.}~\bibnamefont
  {Murgia}}, \bibinfo {author} {\bibfnamefont {V.}~\bibnamefont {Iršič}}, \
  and\ \bibinfo {author} {\bibfnamefont {M.}~\bibnamefont {Viel}},\ }\href
  {\doibase 10.1103/PhysRevD.98.083540} {\bibfield  {journal} {\bibinfo
  {journal} {Phys. Rev.}\ }\textbf {\bibinfo {volume} {D98}},\ \bibinfo {pages}
  {083540} (\bibinfo {year} {2018})},\ \Eprint
  {http://arxiv.org/abs/1806.08371} {arXiv:1806.08371 [astro-ph.CO]}
  \BibitemShut {NoStop}%
\bibitem [{\citenamefont {Anderhalden}\ \emph {et~al.}(2013)\citenamefont
  {Anderhalden}, \citenamefont {Schneider}, \citenamefont {Maccio},
  \citenamefont {Diemand},\ and\ \citenamefont {Bertone}}]{Anderhalden:2012jc}%
  \BibitemOpen
  \bibfield  {author} {\bibinfo {author} {\bibfnamefont {D.}~\bibnamefont
  {Anderhalden}}, \bibinfo {author} {\bibfnamefont {A.}~\bibnamefont
  {Schneider}}, \bibinfo {author} {\bibfnamefont {A.~V.}\ \bibnamefont
  {Maccio}}, \bibinfo {author} {\bibfnamefont {J.}~\bibnamefont {Diemand}}, \
  and\ \bibinfo {author} {\bibfnamefont {G.}~\bibnamefont {Bertone}},\ }\href
  {\doibase 10.1088/1475-7516/2013/03/014} {\bibfield  {journal} {\bibinfo
  {journal} {JCAP}\ }\textbf {\bibinfo {volume} {1303}},\ \bibinfo {pages}
  {014} (\bibinfo {year} {2013})},\ \Eprint {http://arxiv.org/abs/1212.2967}
  {arXiv:1212.2967 [astro-ph.CO]} \BibitemShut {NoStop}%
\bibitem [{\citenamefont {Garzilli}\ \emph {et~al.}(2017)\citenamefont
  {Garzilli}, \citenamefont {Boyarsky},\ and\ \citenamefont
  {Ruchayskiy}}]{Garzilli:2015iwa}%
  \BibitemOpen
  \bibfield  {author} {\bibinfo {author} {\bibfnamefont {A.}~\bibnamefont
  {Garzilli}}, \bibinfo {author} {\bibfnamefont {A.}~\bibnamefont {Boyarsky}},
  \ and\ \bibinfo {author} {\bibfnamefont {O.}~\bibnamefont {Ruchayskiy}},\
  }\href {\doibase 10.1016/j.physletb.2017.08.022} {\bibfield  {journal}
  {\bibinfo  {journal} {Phys. Lett.}\ }\textbf {\bibinfo {volume} {B773}},\
  \bibinfo {pages} {258} (\bibinfo {year} {2017})},\ \Eprint
  {http://arxiv.org/abs/1510.07006} {arXiv:1510.07006 [astro-ph.CO]}
  \BibitemShut {NoStop}%
\bibitem [{\citenamefont {Garzilli}\ \emph {et~al.}(2018)\citenamefont
  {Garzilli}, \citenamefont {Magalich}, \citenamefont {Theuns}, \citenamefont
  {Frenk}, \citenamefont {Weniger}, \citenamefont {Ruchayskiy},\ and\
  \citenamefont {Boyarsky}}]{Garzilli:2018jqh}%
  \BibitemOpen
  \bibfield  {author} {\bibinfo {author} {\bibfnamefont {A.}~\bibnamefont
  {Garzilli}}, \bibinfo {author} {\bibfnamefont {A.}~\bibnamefont {Magalich}},
  \bibinfo {author} {\bibfnamefont {T.}~\bibnamefont {Theuns}}, \bibinfo
  {author} {\bibfnamefont {C.~S.}\ \bibnamefont {Frenk}}, \bibinfo {author}
  {\bibfnamefont {C.}~\bibnamefont {Weniger}}, \bibinfo {author} {\bibfnamefont
  {O.}~\bibnamefont {Ruchayskiy}}, \ and\ \bibinfo {author} {\bibfnamefont
  {A.}~\bibnamefont {Boyarsky}},\ }\href@noop {} {\  (\bibinfo {year}
  {2018})},\ \Eprint {http://arxiv.org/abs/1809.06585} {arXiv:1809.06585
  [astro-ph.CO]} \BibitemShut {NoStop}%
\bibitem [{\citenamefont {Anderhalden}\ \emph {et~al.}(2012)\citenamefont
  {Anderhalden}, \citenamefont {Diemand}, \citenamefont {Bertone},
  \citenamefont {Maccio},\ and\ \citenamefont
  {Schneider}}]{Anderhalden:2012qt}%
  \BibitemOpen
  \bibfield  {author} {\bibinfo {author} {\bibfnamefont {D.}~\bibnamefont
  {Anderhalden}}, \bibinfo {author} {\bibfnamefont {J.}~\bibnamefont
  {Diemand}}, \bibinfo {author} {\bibfnamefont {G.}~\bibnamefont {Bertone}},
  \bibinfo {author} {\bibfnamefont {A.~V.}\ \bibnamefont {Maccio}}, \ and\
  \bibinfo {author} {\bibfnamefont {A.}~\bibnamefont {Schneider}},\ }\href
  {\doibase 10.1088/1475-7516/2012/10/047} {\bibfield  {journal} {\bibinfo
  {journal} {JCAP}\ }\textbf {\bibinfo {volume} {1210}},\ \bibinfo {pages}
  {047} (\bibinfo {year} {2012})},\ \Eprint {http://arxiv.org/abs/1206.3788}
  {arXiv:1206.3788 [astro-ph.CO]} \BibitemShut {NoStop}%
\bibitem [{\citenamefont {Gelmini}\ \emph {et~al.}(2004)\citenamefont
  {Gelmini}, \citenamefont {Palomares-Ruiz},\ and\ \citenamefont
  {Pascoli}}]{Gelmini:2004ah}%
  \BibitemOpen
  \bibfield  {author} {\bibinfo {author} {\bibfnamefont {G.}~\bibnamefont
  {Gelmini}}, \bibinfo {author} {\bibfnamefont {S.}~\bibnamefont
  {Palomares-Ruiz}}, \ and\ \bibinfo {author} {\bibfnamefont {S.}~\bibnamefont
  {Pascoli}},\ }\href@noop {} {\bibfield  {journal} {\bibinfo  {journal} {Phys.
  Rev. Lett.}\ }\textbf {\bibinfo {volume} {93}},\ \bibinfo {pages} {081302}
  (\bibinfo {year} {2004})},\ \Eprint {http://arxiv.org/abs/astro-ph/0403323}
  {astro-ph/0403323} \BibitemShut {NoStop}%
\bibitem [{\citenamefont {Boyanovsky}(2008)}]{Boyanovsky:2007ba}%
  \BibitemOpen
  \bibfield  {author} {\bibinfo {author} {\bibfnamefont {D.}~\bibnamefont
  {Boyanovsky}},\ }\href {\doibase 10.1103/PhysRevD.77.023528} {\bibfield
  {journal} {\bibinfo  {journal} {Phys. Rev.}\ }\textbf {\bibinfo {volume}
  {D77}},\ \bibinfo {pages} {023528} (\bibinfo {year} {2008})},\ \Eprint
  {http://arxiv.org/abs/0711.0470} {arXiv:0711.0470 [astro-ph]} \BibitemShut
  {NoStop}%
\bibitem [{\citenamefont {Petraki}(2008)}]{Petraki:2008ef}%
  \BibitemOpen
  \bibfield  {author} {\bibinfo {author} {\bibfnamefont {K.}~\bibnamefont
  {Petraki}},\ }\href {\doibase 10.1103/PhysRevD.77.105004} {\bibfield
  {journal} {\bibinfo  {journal} {Phys. Rev.}\ }\textbf {\bibinfo {volume}
  {D77}},\ \bibinfo {pages} {105004} (\bibinfo {year} {2008})},\ \Eprint
  {http://arxiv.org/abs/arXiv:0801.3470} {arXiv:arXiv:0801.3470 [hep-ph]}
  \BibitemShut {NoStop}%
\bibitem [{\citenamefont {Boyarsky}\ \emph {et~al.}(2009)\citenamefont
  {Boyarsky}, \citenamefont {Lesgourgues}, \citenamefont {Ruchayskiy},\ and\
  \citenamefont {Viel}}]{Boyarsky:2008xj}%
  \BibitemOpen
  \bibfield  {author} {\bibinfo {author} {\bibfnamefont {A.}~\bibnamefont
  {Boyarsky}}, \bibinfo {author} {\bibfnamefont {J.}~\bibnamefont
  {Lesgourgues}}, \bibinfo {author} {\bibfnamefont {O.}~\bibnamefont
  {Ruchayskiy}}, \ and\ \bibinfo {author} {\bibfnamefont {M.}~\bibnamefont
  {Viel}},\ }\href {\doibase 10.1088/1475-7516/2009/05/012} {\bibfield
  {journal} {\bibinfo  {journal} {JCAP}\ }\textbf {\bibinfo {volume} {0905}},\
  \bibinfo {pages} {012} (\bibinfo {year} {2009})},\ \Eprint
  {http://arxiv.org/abs/0812.0010} {arXiv:0812.0010 [astro-ph]} \BibitemShut
  {NoStop}%
\bibitem [{\citenamefont {Wyman}\ \emph {et~al.}(2014)\citenamefont {Wyman},
  \citenamefont {Rudd}, \citenamefont {Vanderveld},\ and\ \citenamefont
  {Hu}}]{Wyman:2013lza}%
  \BibitemOpen
  \bibfield  {author} {\bibinfo {author} {\bibfnamefont {M.}~\bibnamefont
  {Wyman}}, \bibinfo {author} {\bibfnamefont {D.~H.}\ \bibnamefont {Rudd}},
  \bibinfo {author} {\bibfnamefont {R.~A.}\ \bibnamefont {Vanderveld}}, \ and\
  \bibinfo {author} {\bibfnamefont {W.}~\bibnamefont {Hu}},\ }\href {\doibase
  10.1103/PhysRevLett.112.051302} {\bibfield  {journal} {\bibinfo  {journal}
  {Phys. Rev. Lett.}\ }\textbf {\bibinfo {volume} {112}},\ \bibinfo {pages}
  {051302} (\bibinfo {year} {2014})},\ \Eprint {http://arxiv.org/abs/1307.7715}
  {arXiv:1307.7715 [astro-ph.CO]} \BibitemShut {NoStop}%
\bibitem [{\citenamefont {Battye}\ and\ \citenamefont
  {Moss}(2014)}]{Battye:2013xqa}%
  \BibitemOpen
  \bibfield  {author} {\bibinfo {author} {\bibfnamefont {R.~A.}\ \bibnamefont
  {Battye}}\ and\ \bibinfo {author} {\bibfnamefont {A.}~\bibnamefont {Moss}},\
  }\href {\doibase 10.1103/PhysRevLett.112.051303} {\bibfield  {journal}
  {\bibinfo  {journal} {Phys. Rev. Lett.}\ }\textbf {\bibinfo {volume} {112}},\
  \bibinfo {pages} {051303} (\bibinfo {year} {2014})},\ \Eprint
  {http://arxiv.org/abs/1308.5870} {arXiv:1308.5870 [astro-ph.CO]} \BibitemShut
  {NoStop}%
\bibitem [{\citenamefont {Dvorkin}\ \emph {et~al.}(2014)\citenamefont
  {Dvorkin}, \citenamefont {Wyman}, \citenamefont {Rudd},\ and\ \citenamefont
  {Hu}}]{Dvorkin:2014lea}%
  \BibitemOpen
  \bibfield  {author} {\bibinfo {author} {\bibfnamefont {C.}~\bibnamefont
  {Dvorkin}}, \bibinfo {author} {\bibfnamefont {M.}~\bibnamefont {Wyman}},
  \bibinfo {author} {\bibfnamefont {D.~H.}\ \bibnamefont {Rudd}}, \ and\
  \bibinfo {author} {\bibfnamefont {W.}~\bibnamefont {Hu}},\ }\href {\doibase
  10.1103/PhysRevD.90.083503} {\bibfield  {journal} {\bibinfo  {journal} {Phys.
  Rev.}\ }\textbf {\bibinfo {volume} {D90}},\ \bibinfo {pages} {083503}
  (\bibinfo {year} {2014})},\ \Eprint {http://arxiv.org/abs/1403.8049}
  {arXiv:1403.8049 [astro-ph.CO]} \BibitemShut {NoStop}%
\bibitem [{\citenamefont {Beutler}\ \emph {et~al.}(2014)\citenamefont {Beutler}
  \emph {et~al.}}]{Beutler:2014yhv}%
  \BibitemOpen
  \bibfield  {author} {\bibinfo {author} {\bibfnamefont {F.}~\bibnamefont
  {Beutler}} \emph {et~al.} (\bibinfo {collaboration} {BOSS}),\ }\href
  {\doibase 10.1093/mnras/stu1702} {\bibfield  {journal} {\bibinfo  {journal}
  {Mon. Not. Roy. Astron. Soc.}\ }\textbf {\bibinfo {volume} {444}},\ \bibinfo
  {pages} {3501} (\bibinfo {year} {2014})},\ \Eprint
  {http://arxiv.org/abs/1403.4599} {arXiv:1403.4599 [astro-ph.CO]} \BibitemShut
  {NoStop}%
\bibitem [{\citenamefont {Canac}\ \emph {et~al.}(2016)\citenamefont {Canac},
  \citenamefont {Aslanyan}, \citenamefont {Abazajian}, \citenamefont
  {Easther},\ and\ \citenamefont {Price}}]{Canac:2016smv}%
  \BibitemOpen
  \bibfield  {author} {\bibinfo {author} {\bibfnamefont {N.}~\bibnamefont
  {Canac}}, \bibinfo {author} {\bibfnamefont {G.}~\bibnamefont {Aslanyan}},
  \bibinfo {author} {\bibfnamefont {K.~N.}\ \bibnamefont {Abazajian}}, \bibinfo
  {author} {\bibfnamefont {R.}~\bibnamefont {Easther}}, \ and\ \bibinfo
  {author} {\bibfnamefont {L.~C.}\ \bibnamefont {Price}},\ }\href {\doibase
  10.1088/1475-7516/2016/09/022} {\bibfield  {journal} {\bibinfo  {journal}
  {JCAP}\ }\textbf {\bibinfo {volume} {1609}},\ \bibinfo {pages} {022}
  (\bibinfo {year} {2016})},\ \Eprint {http://arxiv.org/abs/1606.03057}
  {arXiv:1606.03057 [astro-ph.CO]} \BibitemShut {NoStop}%
\bibitem [{\citenamefont {Mertens}\ \emph {et~al.}(2015)\citenamefont
  {Mertens}, \citenamefont {Lasserre}, \citenamefont {Groh}, \citenamefont
  {Drexlin}, \citenamefont {Glueck}, \citenamefont {Huber}, \citenamefont
  {Poon}, \citenamefont {Steidl}, \citenamefont {Steinbrink},\ and\
  \citenamefont {Weinheimer}}]{Mertens:2014nha}%
  \BibitemOpen
  \bibfield  {author} {\bibinfo {author} {\bibfnamefont {S.}~\bibnamefont
  {Mertens}}, \bibinfo {author} {\bibfnamefont {T.}~\bibnamefont {Lasserre}},
  \bibinfo {author} {\bibfnamefont {S.}~\bibnamefont {Groh}}, \bibinfo {author}
  {\bibfnamefont {G.}~\bibnamefont {Drexlin}}, \bibinfo {author} {\bibfnamefont
  {F.}~\bibnamefont {Glueck}}, \bibinfo {author} {\bibfnamefont
  {A.}~\bibnamefont {Huber}}, \bibinfo {author} {\bibfnamefont {A.~W.~P.}\
  \bibnamefont {Poon}}, \bibinfo {author} {\bibfnamefont {M.}~\bibnamefont
  {Steidl}}, \bibinfo {author} {\bibfnamefont {N.}~\bibnamefont {Steinbrink}},
  \ and\ \bibinfo {author} {\bibfnamefont {C.}~\bibnamefont {Weinheimer}},\
  }\href {\doibase 10.1088/1475-7516/2015/02/020} {\bibfield  {journal}
  {\bibinfo  {journal} {JCAP}\ }\textbf {\bibinfo {volume} {1502}},\ \bibinfo
  {pages} {020} (\bibinfo {year} {2015})},\ \Eprint
  {http://arxiv.org/abs/1409.0920} {arXiv:1409.0920 [physics.ins-det]}
  \BibitemShut {NoStop}%
\bibitem [{\citenamefont {{Jelinsky}}\ \emph {et~al.}(1995)\citenamefont
  {{Jelinsky}}, \citenamefont {{Vallerga}},\ and\ \citenamefont
  {{Edelstein}}}]{1995ApJ...442..653J}%
  \BibitemOpen
  \bibfield  {author} {\bibinfo {author} {\bibfnamefont {P.}~\bibnamefont
  {{Jelinsky}}}, \bibinfo {author} {\bibfnamefont {J.~V.}\ \bibnamefont
  {{Vallerga}}}, \ and\ \bibinfo {author} {\bibfnamefont {J.}~\bibnamefont
  {{Edelstein}}},\ }\href {\doibase 10.1086/175469} {\bibfield  {journal}
  {\bibinfo  {journal} {\apj}\ }\textbf {\bibinfo {volume} {442}},\ \bibinfo
  {pages} {653} (\bibinfo {year} {1995})}\BibitemShut {NoStop}%
\bibitem [{\citenamefont {Iršič}\ \emph {et~al.}(2017)\citenamefont {Iršič}
  \emph {et~al.}}]{Irsic:2017ixq}%
  \BibitemOpen
  \bibfield  {author} {\bibinfo {author} {\bibfnamefont {V.}~\bibnamefont
  {Iršič}} \emph {et~al.},\ }\href {\doibase 10.1103/PhysRevD.96.023522}
  {\bibfield  {journal} {\bibinfo  {journal} {Phys. Rev.}\ }\textbf {\bibinfo
  {volume} {D96}},\ \bibinfo {pages} {023522} (\bibinfo {year} {2017})},\
  \Eprint {http://arxiv.org/abs/1702.01764} {arXiv:1702.01764 [astro-ph.CO]}
  \BibitemShut {NoStop}%
\end{thebibliography}%
\end{document}